\documentclass[twocolumn]{aastex631}

\begin{document}

\title{On the Planetary Theory of Everything}

\author{J.J. Charfman Jr.}
\affiliation{Department of Astronomy and Steward Observatory, The University of Arizona, Tucson, AZ 85721, USA}
\affiliation{Lunar and Planetary Laboratory, The University of Arizona, Tucson, AZ 85721, USA}

\author{M. M. M.}
\affiliation{Department of Astronomy and Steward Observatory, The University of Arizona, Tucson, AZ 85721, USA}

\author{J. Dietrich}
\affiliation{Department of Astronomy and Steward Observatory, The University of Arizona, Tucson, AZ 85721, USA}

\author{N. T. Schragal}
\affiliation{Department of Astronomy and Steward Observatory, The University of Arizona, Tucson, AZ 85721, USA}

\author{A. M. Avsar}
\affiliation{Lunar and Planetary Laboratory, The University of Arizona, Tucson, AZ 85721, USA}




\begin{abstract}
Here, we present a simple solution to problems that have plagued (extra)``galactic" astronomers and cosmologists over the last century. We show that ``galaxy" formation, dark matter, and the tension in the expansion of the universe can all be explained by the natural behaviors of an overwhelmingly large population of exoplanets throughout the universe. Some of these ideas have started to be proposed in the literature, and we commend these pioneers revolutionizing our understanding of astrophysics. Furthermore, we assert that, since planets are obviously the ubiquitous answer to every current question that can be posed by astronomers, planetary science must then be the basis for all science, and therefore that all current funding for science be reserved for (exo)planetary science - we happily welcome all astronomers and other scientists.
\end{abstract}



\section{Introduction} \label{sec:intro}

It has come to our attention that a regrettably large number of astronomers do not believe that the existence of planets outside our solar system can be proven (Woodrum, Hviding, Amaro, and Chamberlain \citeyear{2023arXiv}). These astronomers must have their head in the interstellar clouds, since they cannot see the overwhelming evidence that exoplanets are everywhere. By number, exoplanets are the most common objects of their size or larger in the universe by at least an order of magnitude. We have evidence for an average of more than one planet orbiting each star \citep[e.g.,][]{2021ARA&A..59..291Z}, and this does not include the expected number of free-floating planets that did not form around host stars or have been ejected from their original systems \citep{2021MNRAS.505.5584M}. Estimates predict upwards of one hundred thousand free-floating planets per star in the universe - planets that either possibly formed on their own in mini-collapses of very locally concentrated matter, or more likely were kicked out of their nascent protoplanetary disk by a bigger badder neighborhood bully planet \citep[see e.g,][]{2012MNRAS.423.1856S}.

Due to their overwhelming ubiquity, we instead propose that planets are the solution to the current enigmas of astronomy. As their presence and importance to the field has been proven time and again over the past three decades, we must now begin to expand our planetary horizons and test to see how much can be explained by a universal ``planetary theory of everything." There are many questions about the universe that are quickly waved away by some explanation of ``dark" whatever, simply because no one has deigned to believe the answer could just be planets.

This paper is organized as follows. In Section 2 we focus on the similarities of ``galaxies" with protoplanetary clouds and disks and how therefore ``galaxies" are simply planetary systems forming on a large cosmic scale. We discuss the ``planets as dark matter" theory and add our own discussion in Section 3. In Section 4 we show that planets can help resolve the Hubble tension between early-universe Planck CMB measurements and recent-universe SN measurements of the Hubble constant. Finally, we conclude that planetary science now encompasses all of astronomy and provide a reasonable statement on what that means for future astronomical funding in Section 5.

\section{``Galaxies" Are Cosmic Planetary Systems}

\subsection{Morphology, and the return of the tuning fork}

``Galaxy" morphology typically splits ``galaxies" into two primary categories: spiral and elliptical. Early in the history of extra-``galactic" astronomy, it was theorized that these two morphological classes were the beginning and end states of an evolutionary sequence commonly referred to as the ``Hubble tuning fork". In this sequence, all ``galaxies" begin as diffuse and bulbous elliptical ``galaxies", and over time coalesce into a spiral structure. However, as a popular theory of these objects being ``galaxies" took hold, successive theories rejected this evolutionary sequence. In contrast, this evolutionary sequence from amorphism to order precisely matches how planetary systems form and evolve. The observed morphology of so-called ``galaxies" is a natural result of planetary theory.

\begin{figure*}[ht]
    \centering
    \includegraphics[width=\textwidth]{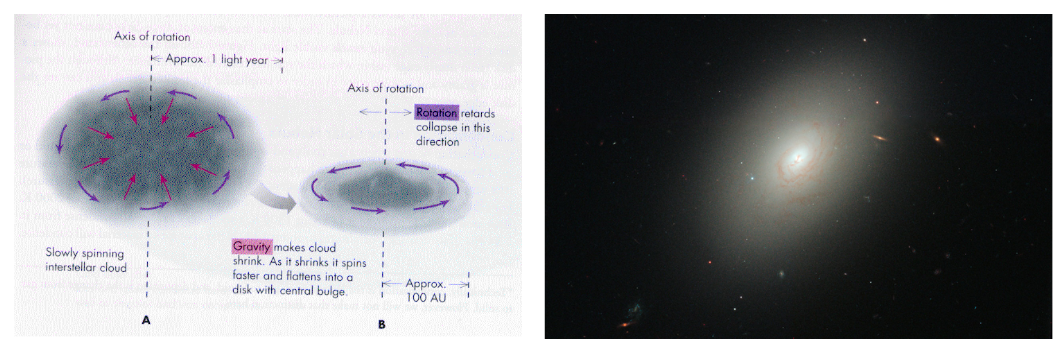}
    \caption{
    \textbf{Left:} a protoplanetary cloud evolving to form a protoplanetary disk \textbf{Right:} an ``elliptical galaxy". The ``galaxy" shows the same nebulous elliptical structure with a concentrated center and slowly radially decreasing brightness profile as the protoplanetary cloud. \textbf{Thus, the ``galaxy" is simply a protoplanetary cloud before its evolution into a protoplanetary disk.}
    \\
    Credit http://burro.case.edu/Academics/Astr221/SolarSys/Formation/starform.html and\\ https://esahubble.org/wordbank/elliptical-galaxy/}
    \label{fig:cloud_v_elliptical}
\end{figure*}

Protoplanetary disks form via the collapse of clouds of gas. The initial cloud is amorphous, which explains the observed shape of so-called ``elliptical galaxies". Figure~\ref{fig:cloud_v_elliptical} compares a sketch of a protoplanetary cloud to the observed object NGC 4150. The ovular, egg-like shape of NGC 4150 matches the shape of a protoplanetary cloud. Therefore, the best explanation is that this so-called ``elliptical galaxy" is actually a protoplanetary cloud which is likely in the process of collapsing into a cosmic-scale protoplanetary disk, from which a plethora of planetary systems will be born. This explanation is strongly consistent with the Hubble tuning fork, as a Yakov-Smirnov B-S test of separating populations or models \citep[see e.g.,][even if their main finding has since been challenged]{2002astro.ph..4041C} cannot distinguish between the evolution of protoplanetary clouds/disks and the Hubble tuning fork for ``galaxy" morphology at a significant level.

\begin{figure*}[ht]
    \centering
    \includegraphics[width=\textwidth]{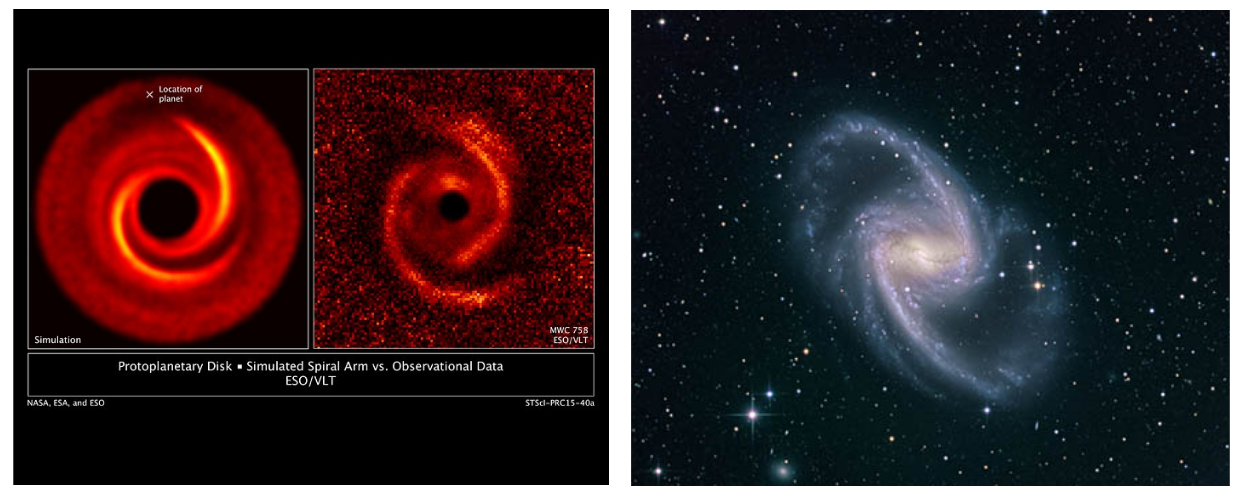}
    \caption{
    \textbf{Left:} a simulation of a protoplanetary disk, compared to a real protoplanetary disk system MWC 758 observed by VLT. \textbf{Right:} an image of NGC 1365, here referred to as a ``spiral galaxy". Note that the morphology of this supposed ``galaxy", complete with spiral arms, a potential barred center, and overall brightness of the armed structure, precisely matches both the simulated and observed protoplanetary disk. \textbf{It is thus clear that NGC 1365 is instead a protoplanetary disk.}
    \\
    Credit https://esahubble.org/images/opo1540a/ and http://www.astropixels.com/galaxies/NGC1365-CDK21-C01.html}
    \label{fig:disk_v_spiral}
\end{figure*}

The next stage of planet formation includes the congolometration and accretion of solids into protoplanets within the protoplanetary disk. Once large enough, these protoplanets interact with the disk and create structures like spirals. Figure~\ref{fig:disk_v_spiral} shows the result of a simulation of a protoplanet embedded within its disk. Spirals and density waves naturally appear, and even provide observable evidence for that planet. These planet-induced structures are precisely what we see from so-called ``spiral galaxies" which, as shown in Figure~\ref{fig:disk_v_spiral}, exhibit the same exact spiral arms and density fluctuations predicted in the protoplanetary disk simulation. Therefore, these ``spiral galaxies" must be large-scale disks actively forming a sizeable population of planets. This explanation is fully consistent with these spiral systems following the aforementioned elliptical systems in the evolutionary tuning fork. 

\subsection{Planetary Formation Scales}

Since we have now shown that ``galaxies" themselves are rather protoplanetary clouds and disks, we must address the scales of planetary formation. Local planetary systems form around stars, which happens on length scales of AU and timescales of millions of years. On the other hand, cosmic-scale planetary systems that form around the central core super-star at the center of these ``galaxies" have a greater length scale of kiloparsecs, and their timescale must correspondingly be larger by a similar factor.

With the distance of 1 AU when expressed in centimeters only a factor of $\sim2$ different from 1 Myr when expressed in seconds, we can assume that this scaling for cosmic protoplanetary systems must roughly be of a similar magnitude. Thus, a length scale of 1 kpc in cm would then roughly correspond to a timescale of 100 trillion years, which is much longer than the age of the universe. Therefore, at just 14 billion years we are only seeing a snapshot of a few early disk-forming cosmic planetary systems, along with some still-nebulous cosmic protoplanetary clouds that have yet to start their disk formation phase.

\section{Planets as Dark Matter}

Since ``galactic" astronomers don't believe in planets, they have been missing the most obvious candidate for dark matter. The invisible and almost purely gravitational effects of dark matter on ``galaxies" and ``galaxy clusters" can easily be explained by bunches of the aforementioned free-floating planets \citep[][]{2012MNRAS.423.1856S}, or even dark matter exoplanets that have been theorized to exist \citep[][]{2023arXiv230312129B}.

Dark matter is theorized to have different forms separated by the initial budget of kinetic energy/temperature, imaginatively named ``cold dark matter" and ``warm/hot dark matter". Cold dark matter (CDM) is expected to comprise a majority of the dark matter in the universe, forming a halo around individual ``galaxies" that speeds up their differential rotation curves, as well as interacting with visible matter within the intracluster medium of ``galaxy clusters" and along filaments of large-scale ``galactic" structure in the universe.

\subsection{Dark Matter Deficient ``Galaxies"}

A recent problem that has emerged in ``galaxy" formation theories is the existence of dark matter deficient ``galaxies": diffuse satellite systems with little to no dark matter halo surrounding them \citep[see e.g.,][]{2018Natur.555..629V}.  Many teams of theorists have set out to solve the problem of dark matter deficient ``galaxies" by running complicated and computationally intensive cosmological simulations to explain how these ``galaxies" can exist \citep[e.g.,][]{2022NatAs...6..496M}. Since these theorists most likely do not believe in the existence of exoplanets, they have been missing the simple solution all along. 

We argue that these ``galaxies" are not dark matter deficient, but planet deficient. To prove this argument, we introduce the Charfman-Avsar relation, which is as follows: 
\begin{equation}
    \text{Less Planets} = \text{Less Mass}
\end{equation}
Other groups have tried to explain dark matter deficiencies through complex ``galactic" evolution, involving tidal forces and ``galactic" mergers. Our less complicated proposition would be able to explain all observations through a single mechanism, reducing computational time for cosmological simulations. Although we expect there to be pushback from the larger community, we assure this proposition is airtight and should be adopted immediately. 

\subsection{MACHO Cold Dark Matter}

Previous studies have looked into the presence of planetary-mass non-self-luminous objects floating in the outer reaches of ``galaxies" that comprise this material that only interacts gravitationally with the rest of the visible matter in the universe. These ``MAssive Compact Halo Objects" \citep[MACHOs, e.g.,][and many sources afterwards]{1990Natur.345..478C, 1993NYASA.688..390G}. Perhaps \textit{not} coincidentally, the field of exoplanets and the field of MACHO-dominated dark matter evolved contemporaneously in the early 1990s, suggesting a common basic ideology.

Much additional work has since come out to ``disprove" the MACHO dark matter theory, mostly in the form of microlensing surveys \citep[e.g.,][]{2000ApJ...542..281A, 2007A&A...469..387T}. These results claim that a fully MACHO dark matter halo is inconsistent with their results, and that a fractional MACHO halo may be more likely but still requires an additional component currently unknown to astronomy. However, microlensing is inherently biased with requiring a dense stellar background, and even observing stars in front of the Milky Way bulge for over a decade has produced $<200$ microlensing planets bound around their host stars (per the NASA Exoplanet Archive as of the publication of this manuscript), plus a few microlensing free-floating planets \citep[][]{2021MNRAS.505.5584M}. Thus, we claim (contrary to any previous evidence to counter this) that observing a small fraction of the Milky Way halo towards the Magellanic Clouds simply does not provide a dense enough stellar background to detect microlensing free-floating planets in the ``galactic" halo to a large enough degree.

\subsection{Dark Matter Planets}

While we have definitively proven that dark matter is just planets, we can also look into the hypothesis of different planet-sized objects made of dark matter that are not themselves planets. \citet[][]{2023arXiv230312129B} state that if dark matter exoplanets exist, they would mostly be indistinguishable from regular matter planets via the transit method unless they are large or of low opacity, neither of which is likely with current theories. However, since this requires a more complicated theory than dark matter simply being regular planets, especially since the difference is negligible for most of the parameter space, we reject that hypothesis.

\begin{figure*}[ht]
    \centering
    \includegraphics[width=\textwidth]{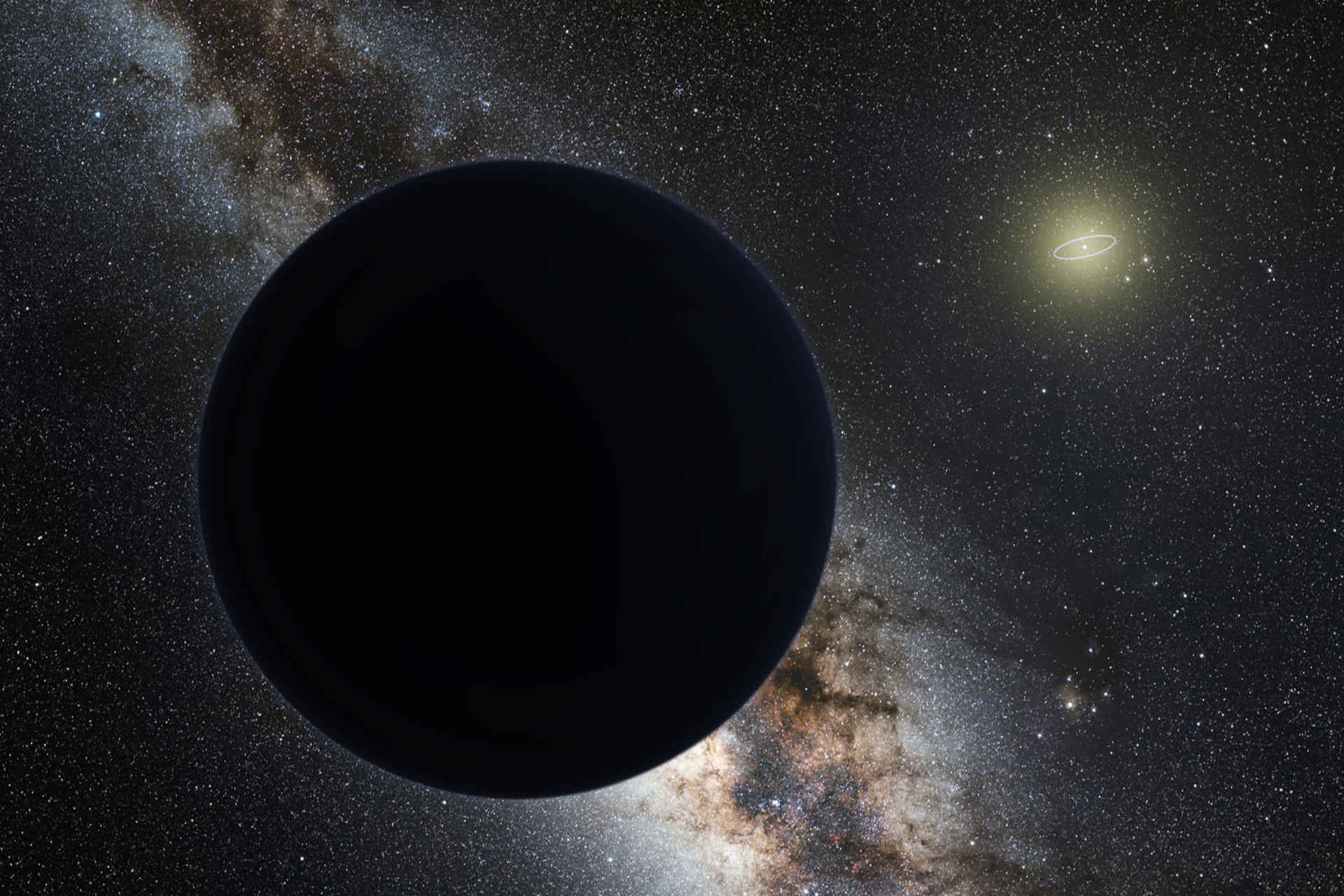}
    \caption{An artist's impression of an unseen planet in the outskirts of our solar system - Planet Nine, MACHO, dark matter, etc. \textbf{It's all the same thing.} Image Credit: nagualdesign, Tom Ruen, and ESO}
    \label{fig:P9DM}
\end{figure*}

\citet[][]{2022arXiv220317075P} declare that, although we have a good concept of the planetary components of our solar system, we could still discover an additional one by methods similar to those used to currently support the existence of Planet Nine \citep[see e.g.,][]{2016ApJ...824L..23B}. Indeed, a non-self-luminous planetary body hiding inside our own solar system unable to be detected by anything other than its affect on the orbits of other bodies is very similar to what we see with the effects of dark matter across ``galaxies". Therefore, with an example from an artist's rendition seen in Figure~\ref{fig:P9DM}, we unequivocally state that dark matter is just unseen planets gravitationally affecting planetary systems on both the stellar and cosmic scale.

There is also the Warm Dark Matter (WDM) component that was previously believed to simply be the same as cold dark matter, but with a relativistic initial energy bucket. If this is true, it might have enough energy to glow thermally even though it is dark optically. \citet[]{2022arXiv220316563L} found that lava matches many of the same observational signatures as WDM and is therefore a strong candidate to explain WDM. Thus, rocky planets that are cold and dark on the outside yet volcanically active on the inside could be both the CDM we see everywhere as well as the source of WDM that is less prevalent across the universe.

\section{Resolving the Hubble Tension with Planets}

One of the fundamental cosmological parameters of which we know surprisingly little about is the Hubble constant of the expanding universe. The original measurements by Edwin Hubble placed the value somewhere around 500 km/s/Mpc \citep[][and note how he may have been ahead of the time with calling these ``galaxies" outside of the Milky Way as ``nebulae"]{1929PNAS...15..168H}, and as recently as the mid-1990s the actual value was still debated to be between 50 and 100 km/s/Mpc \citep[][]{1996PASP..108.1065B}. However, recent data from Planck measuring the CMB anisotropies gave a value of $67.66\pm0.42$ \citep[][]{2020A&A...641A...6P}, whereas distance ladder measurements using Cepheid variables and Type Ia SN provide a value of $73.04\pm1.04$ \citep[][]{2022ApJ...934L...7R}, which are discrepant at 5$\sigma$.

\begin{figure*}[ht]
    \centering
    \includegraphics[width=\textwidth]{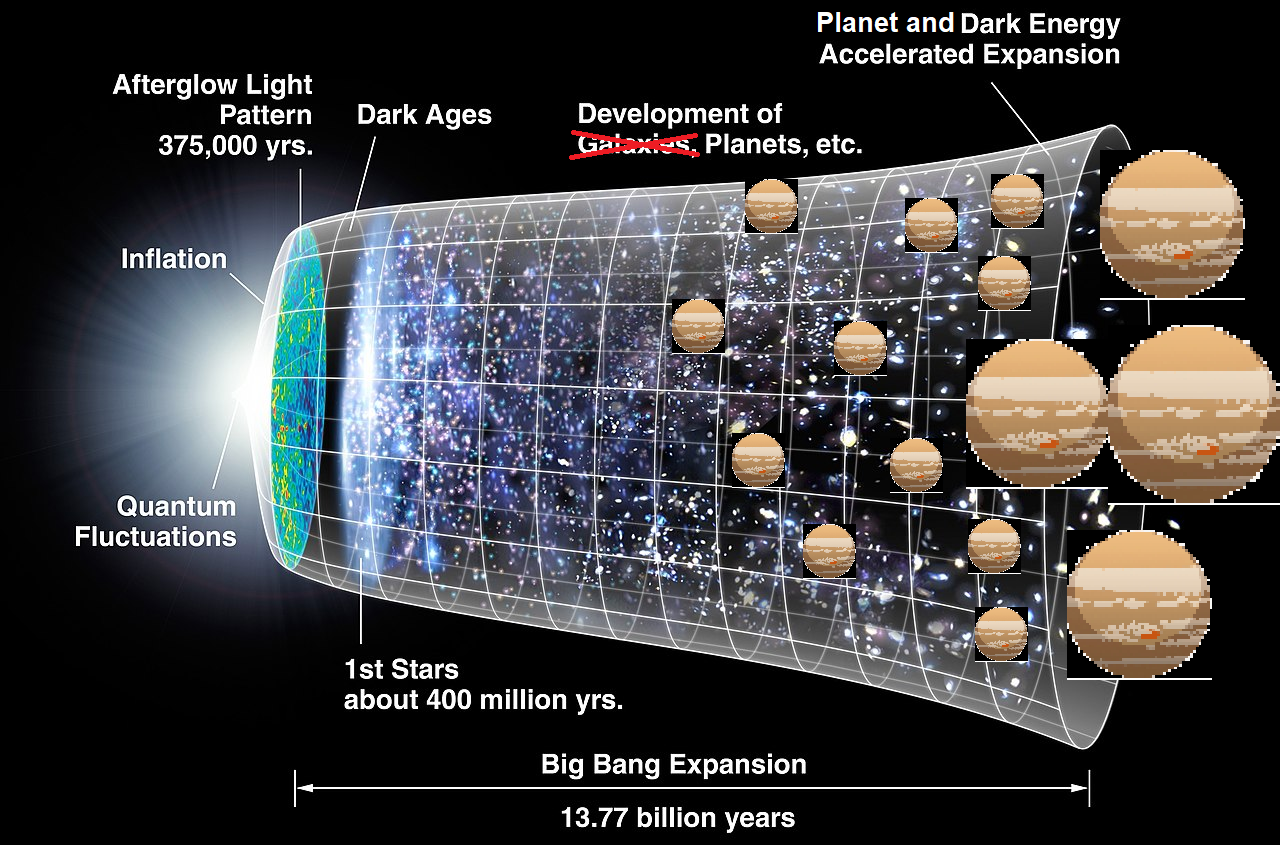}
    \caption{An artist's impression of the expansion of the universe, with the change in the acceleration fueled by the increased number of planets in the recent universe. Original Image Credit: NASA/WMAP}
    \label{fig:expansion}
\end{figure*}

Once again, the obvious solution to this problem is planets, specifically the planet formation timescale (as referenced above in Section 2.2). The Planck measurements of the CMB anisotropies were done in such an early universe that we are sure there were absolutely no planets around at that point, so the presence of planets was not included in determining the Hubble constant, which provides the lower value. However, the distance ladder measurements come from the recent universe, which as we all know is teeming with planets. Therefore, it is obvious that the presence of planets in the recent universe have simply added on another parameter to the Friedman equation for the universe, such that the Hubble constant has increased to its current value. We introduce this as a natural corollary of the Charfman-Avsar relation:
\begin{equation}
    \text{Corollary 1: } \frac{\text{More Planets}}{\text{Unit Time}} \propto \frac{\text{Higher } H_0}{\text{Unit Time}}
\end{equation}
Therefore, both the Planck measurements and the distance ladder measurements can be right in their own observation epochs! See Figure~\ref{fig:expansion} for an artist's impression of this phenomenon.

Another interesting recent measurement on the Hubble constant is from \citet[][]{2022arXiv220316551A}, done in the extreme local and extreme recent universe with the Moon's orbital recession from Earth. While this is a method ingenious in its simplicity with a precise result, we believe it is not as free from systematic biases and errors as they claim. The measurement is only done with one planet and one satellite object, which is hardly representative of the universe as a whole with its untold magnitude of planets, and the measurement is also only corrected for tides without taking into account any other higher-order issues that may occur. We suspect this is the reasoning for the measurement below even the Planck value, even though they do include more planetary bodies than the Planck measurement does.

\section{Summary}

\begin{itemize}
    \item Planets are ubiquitous in the universe and are the most common object known with their own self-gravity.
    \item ``Galaxies" are really just cosmically large planetary systems evolving on timescales longer than the age of the universe.
    \item Dark matter actually does follow the MACHO paradigm, because it is trillions of free-floating non-luminous planets.
    \item The formation of planets as the numerically dominant object in the universe over its age has also caused an increase in the Hubble constant over that same age.
\end{itemize}

We recommend that ``planetary science", hereafter, should just be known as ``science" since we have shown that the planetary aspect is all-encompassing. And since funding for science has been shown to be extremely important and correlated with many different success metrics for universities and institutes of higher study, we assert that this funding must be used as we scientists see fit. We promise that all astronomers and other scientists of previously-branded branches of the field will be welcomed into this new and exciting re-organization of science.

The authors would like to thank B.S. Prince at the Lunar and Planetary Laboratory for his help in making the figures and for his lucid skepticism of modern astronomy, as well as Orion and Luna the cats for interesting non-scientific discussions. This paper was written in remembrance of J.J. Charfman, who will be dearly missed even as their legacy lives on (love you Mapa). This ``research" has made use of the NASA Exoplanet Archive, which is operated by the California Institute of Technology, under contract with the National Aeronautics and Space Administration under the Exoplanet Exploration Program. 

\bibliography{sample631}{}
\bibliographystyle{aasjournal}



\end{document}